\documentclass{ieeeaccess}
\usepackage{cite}
\usepackage{amsmath,amssymb,amsfonts}
\usepackage{algorithmic}
\usepackage{graphicx}
\usepackage{textcomp}
\def\BibTeX{{\rm B\kern-.05em{\sc i\kern-.025em b}\kern-.08em
    T\kern-.1667em\lower.7ex\hbox{E}\kern-.125emX}}
\begin{document}
\doi{-}

\title{Qubit-compatible substrates with superconducting through-silicon vias}
\author{
\uppercase{K. Grigoras}\authorrefmark{1}, 
\uppercase{N. Yurttag{\"u}l\authorrefmark{1}}, 
\uppercase{J.-P. Kaikkonen\authorrefmark{1}},  
\uppercase{E. T. Mannila\authorrefmark{1}}, 
\uppercase{P. Eskelinen\authorrefmark{1}},  
\uppercase{D. P. Lozano\authorrefmark{2}}, 
\uppercase{H.-X. Li\authorrefmark{2}}, 
\uppercase{M. Rommel\authorrefmark{2}}, 
\uppercase{D. Shiri\authorrefmark{2}}, 
\uppercase{N. Tiencken\authorrefmark{1}},  
\uppercase{S. Simbierowicz\authorrefmark{1}},  
\uppercase{A. Ronzani\authorrefmark{1}},  
\uppercase{J. H{\"a}tinen\authorrefmark{1}},  
\uppercase{D. Datta\authorrefmark{1}},  
\uppercase{V. Vesterinen\authorrefmark{1}},  
\uppercase{L. Gr{\"o}nberg\authorrefmark{1}},  
\uppercase{J. Bizn{\'a}rov{\'a}\authorrefmark{2}},  
\uppercase{A. Fadavi Roudsari\authorrefmark{2}},  
\uppercase{S. Kosen\authorrefmark{2}},  
\uppercase{A. Osman\authorrefmark{2}},  
\uppercase{M. Prunnila\authorrefmark{1}}, 
\uppercase{J. Hassel\authorrefmark{1}},  
\uppercase{J. Bylander\authorrefmark{2}},  
\uppercase{J. Govenius\authorrefmark{1}} 
}
\address[1]{VTT Technical Research Centre of Finland Ltd., QTF Centre of Excellence, P.O. Box 1000, FI-02044 VTT Espoo, Finland}
\address[2]{Department of Microtechnology and Nanoscience, Chalmers University of Technology, 412 96 Gothenburg, Sweden}

\markboth
{Grigoras \headeretal: Qubit-compatible substrates with superconducting through-silicon vias}
{Grigoras \headeretal: Qubit-compatible substrates with superconducting through-silicon vias}

\corresp{Corresponding author: E. T. Mannila (email: elsa.mannila@vtt.fi)}

\begin{abstract}
We fabricate and characterize superconducting through-silicon vias and
electrodes suitable for superconducting quantum processors. We measure
internal quality factors of a million for test resonators excited at
single-photon levels, on chips with superconducting vias used to
stitch ground planes on the front and back sides of the chips. This
resonator performance is on par with the state of the art for
silicon-based planar solutions, despite the presence of vias. Via
stitching of ground planes is an important enabling technology for
increasing the physical size of quantum processor chips, and is a
first step toward more complex quantum devices with three-dimensional
integration.
\end{abstract}


\titlepgskip=-15pt

\maketitle

\section{Introduction}

\IEEEPARstart{P}{erformance} of superconducting qubits
has greatly improved since the first demonstrations of quantum coherence,
with dephasing time, in
particular, increasing four orders of magnitude from 20 ns
demonstrated by Chiorescu et al. in 2003
\cite{chiorescu_coherent_2003} to hundreds of microseconds measured recently
\cite{place_new_2021,wang_towards_2022,gordon_environmental_2022,somoroff_millisecond_2021}.
To an extent, this astonishing progress in coherence time has been achieved
by avoiding complexity in fabrication. State-of-the-art superconducting qubits
are typically fabricated using an extremely restricted set of materials, a low
thermal budget, and a minimal number of depositions and lithographic steps.

Besides long coherence times required to achieve high-fidelity single and two-qubit gates,
quantum computers also need to become
sufficiently large to solve useful computing tasks. For example, tens or
hundreds of millions of physical qubits are likely required for factoring
thousand-bit numbers using Shor's algorithm \cite{fowler_surface_2012}
and similar estimates have been given for quantum chemistry applications \cite{reiher_elucidating_2017}.
These estimates assume error correction based on the surface code \cite{bravyi_quantum_1998},
which is currently the most promising approach to quantum
error correction.
One attractive feature of the surface code is that it requires only two-dimensional nearest-neighbor coupling between qubits,
which makes a physical implementation of a large quantum computer more feasible.
Nevertheless, separate control and readout lines still need to address essentially all of
the qubits. Routing the control and readout lines around coherent qubit couplers
necessitates the use of more than a single electrode layer in larger
quantum processors. Consequently, moving to more complex fabrication seems
unavoidable, either monolithically
or by using multichip modules.
Flip-chip bonded modules of two chips connected by superconducting bumps
increase the layer count to two and air bridges further alleviate routing
challenges. These have indeed been used successfully to construct processors
of several dozen qubits
\cite{arute_quantum_2019,wu_strong_2021,jurcevic_demonstration_2021}, although
with coherence times and gate fidelities significantly lower than in planar
\cite{gambetta_investigating_2017,dunsworth_characterization_2017,nersisyan_manufacturing_2019,burnett_decoherence_2019,osman_simplified_2021}
or flip-chip bonded \cite{kosen_building_2022} single- or few-qubit devices.

Superconducting vias compatible with high-coherence qubits are an important
next step toward larger processors. In addition to routing purposes, so-called
via stitching is likely needed to shunt nominally grounded planes in different
layers to control and push up the frequencies of harmful parasitic
microwave modes that become problematic in physically large chips
\cite{wenner_wirebond_2011}. Traditional integrated circuit vias are, however,
optimized for different goals, such as high normal-state conductivity and
reduction of parasitic capacitance,
instead of superconductivity and extremely low microwave loss required for
qubit compatibility. Integrating their fabrication with qubits also poses
challenges related to material compatibilities and the low thermal budget of
aluminum-based qubits. Superconducting vias have long been used for multilayer
wiring in superconducting quantum interference (SQUID) and single-flux quantum
(SFQ) devices \cite{tolpygo_deep_2014, kiviranta_multilayer_2016}, but the
vias are shallow and pass through amorphous dielectric layers with poor
microwave performance.

Yost et al. \cite{yost_solid-state_2020,mallek_fabrication_2021} have on the
other hand demonstrated through-silicon vias (TSVs) that have a relatively high
aspect ratio and high critical currents, and show promise in terms of not
destroying qubit coherence,
as the demonstrated qubit relaxation time
of 12.5 \textmu s \cite{yost_solid-state_2020} and resonator internal quality
factors of $10^{5}$ to $2\times10^{5}$ \cite{mallek_fabrication_2021} were
identified to be limited by factors unrelated to TSVs.
For comparison,
widely-reproduced relaxation times for transmon qubits on silicon substrates
are near 50 \textmu s
\cite{gambetta_investigating_2017,dunsworth_characterization_2017,nersisyan_manufacturing_2019,burnett_decoherence_2019,osman_simplified_2021,kosen_building_2022}%
. In addition, Gordon et al. have reported relaxation times of hundreds of
\textmu s \cite{gordon_environmental_2022}.
Corresponding widely-reproduced
resonator quality factors are roughly one million for typical co-planar
waveguide (CPW) test resonator geometries
\cite{dunsworth_characterization_2017,nersisyan_manufacturing_2019,mcrae_materials_2020,earnest_substrate_2018,burnett_noise_2018}%
,
although this can be exceeded with deep trenching or short-lived oxide removal
treatments
\cite{calusine_analysis_2018,altoe_localization_2020,verjauw_investigation_2021}%
. Resonator quality factor is often used as a diagnostic predictor of qubit
relaxation time for a qubit with electrodes fabricated using the same flow as
the resonators. Others have also fabricated superconducting TSVs but the microwave
performance of those approaches remains to be measured
\cite{jhabvala_kilopixel_2014,vahidpour_superconducting_2017,alfaro-barrantes_highly-conformal_2021}%
. Furthermore, coherence times exceeding 300 \textmu s have recently been
demonstrated for transmon qubits on planar sapphire substrates
\cite{place_new_2021, wang_towards_2022}. However, typical methods of etching
high aspect ratio TSVs, like the Bosch process \cite{laermer_method_1996, wu_high_2010},
are not available for sapphire substrates.

In this article, we report on resonator internal quality factors of roughly a
million measured on chips with TSVs stitching the ground planes on the front
and back. The TSVs and resonators are fabricated on full 150 mm wafers, with a
via-last approach where the first electrode layer is deposited and patterned
before via formation.
The via-last approach helps create a high-quality interface
between the substrate and the critical electrode layer used for the
resonators since the electrode layer is deposited on virgin wafers before other
potentially harmful processing steps.
Furthermore, the tantalum-based electrode layer used here has
relatively low kinetic inductance, similar to commonly used niobium films.
While high kinetic inductance is useful in superinductors, for example, in the
fluxonium shunt inductor \cite{manucharyan_fluxonium:_2009}, kinetic
inductance in electrodes of transmon qubits is not typically desirable. Low
kinetic inductance tends to also yield better parameter control since
geometric inductance is often more accurately reproducible than kinetic inductance,
which tends to be very sensitive to variations in chemical composition and crystal structure.

\section{Device structure}

\Figure[t!](topskip=0pt, botskip=0pt, midskip=0pt){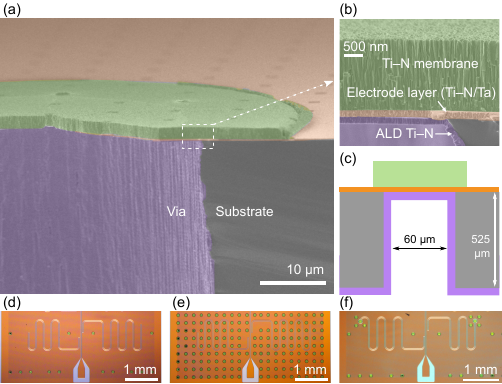}
{(a,b) False-color scanning electron microscope images of a TSV. 
(c) Schematic cross-section of TSV structure (not to scale), with color-coding as in (a,b).
 (d,e,f) Optical
micrographs showing layouts of CPW test resonators with different densities and roles of TSVs (green circles): (d) sparse via stitching, (e) dense via stitching,
(f) TSV-terminated resonators. 
\label{fig:structure}}

Fig. \ref{fig:structure} presents our TSV structure consisting of the main
electrode layer on the front side of the wafer, a hollow via with metallized
walls, a metal membrane covering the via on the front side of the wafer, and a
metallized back side (not visible).
The fabrication process begins with sputtering the main electrode layer, i.e. a bilayer of 15 nm of titanium nitride and 200 nm of tantalum (orange in Fig. \ref{fig:structure}),
on high-resistivity silicon.
For brevity, we refer to this as the
Ta-based electrode layer.
We then pattern the electrode layer using photolithography and
plasma etching.
Next we deposit a sacrificial silicon dioxide layer using plasma-enhanced chemical vapor deposition
and pattern holes in it for the membranes, using photolithography and plasma etching.
We then sputter a 2-\textmu m-thick titanium nitride layer
and pattern it into circular membranes
using photolithography and plasma etching (green in Fig. \ref{fig:structure}).
We choose membrane sputtering parameters that yield relatively low compressive stress of approximately 190 MPa,
as measured on 250-nm-thick reference films.
Next we define the via holes on the back side of the wafer using photolithography
and etch them using the Bosch process.
Finally, we coat the inner walls of the vias and the back side
of the wafer with Ti\textendash N by using
plasma-enhanced atomic layer deposition (ALD, purple in Fig. \ref{fig:structure}),
and remove the sacrificial silicon oxide layer from the front side.
The ALD film thickness is 260 nm, as measured on the back side of the wafer.
As seen in Fig. \ref{fig:structure}(b), the film is noticeably thinner at the other end of the via (ca. 200 nm),
as is typical for plasma-enhanced ALD processes.

The aspect ratio of our TSVs is approximately eight, with a nominal TSV
diameter of 60 \textmu m and substrate thickness of 525 \textmu m. The aspect
ratio is similar to that of Refs.
\cite{yost_solid-state_2020,mallek_fabrication_2021}.
It should be possible to increase the aspect
ratio in the future since the only coating needed inside the vias is produced
by ALD, which is highly conformal compared to most deposition methods.
Furthermore, a smaller via diameter is likely
achievable with thinner wafers, even without increasing aspect ratio.
Smaller-diameter vias
enable increased via density and are likely to lead to increased mechanical robustness of the
metal membranes covering the vias.
Increased via density is likely to be beneficial in the future, when footprint per qubit decreases below current typical values of roughly 0.5 mm$^2$.
Increased mechanical robustness on the other hand improves post-processability.
Currently, the membranes survive typical wafer level handling and processing, but
the suspended parts of the membranes are susceptible to being cleaved off
when the wafers are diced into chips.
The suspended part of the membrane is inconsequential from
the point of view of electrical connectivity, so this
is not a significant issue for the resonator samples
characterized here, but the fragility of the membranes may be an inconvenience
in some applications requiring post processing on diced chips.
Optimizing the stress of the membrane layer to optimally pretension the membrane
could be another future path toward improving mechanical robustness.

Our measurements and results focus on CPW resonators
patterned on chips with TSVs, to demonstrate long relaxation time in the presence of TSVs.
We compare these to planar reference chips with
resonators patterned on Nb, or on the same Ta-based electrode layer as on the TSV
chips. The TSV chips have two to eight CPW resonators coupled to a common
feedline through which transmission is measured (see Fig. \ref{fig:structure}%
). Each resonator acts as a bandstop filter at each of its resonance
frequencies, and thus provides a sensitive probe of microwave loss at those
frequencies, assuming the internal quality factor $Q_{i}$ and coupling quality
factor $Q_{c}$ are of similar order of magnitude. Here, the resonators are
open near the feedline and shorted at the opposite end, with geometry chosen
such that the fundamental $\lambda/4$ resonance frequency varies between 4 GHz and
8 GHz and the coupling quality factor between $2\times10^{4}$ and
$7\times10^{6}$. The width of the CPW center trace is 20 \textmu m and the gap
between the center and ground is 10 \textmu m, as in Ref.
\cite{burnett_noise_2018}. Overetching past the metal layer is small, less
than roughly 50 nm, and no additional trenching is applied. The exact
dimensions of the CPW cross-section play a significant role when making direct
quantitative comparisons since, in extremely low-loss resonators, losses are
generally dominated by material imperfections in thin interface layers between
different materials, and the participation factors of different interfaces are
somewhat geometry dependent
\cite{mcrae_materials_2020,woods_determining_2019,lahtinen_effects_2020,niepce_geometric_2020}.

In terms of density and role of TSVs, we use four types of
layouts: (1) no TSVs and no ground plane on the back, used as reference, (2)
sparse TSVs stitching the ground planes on the front and back, (3) dense TSVs
stiching the ground planes, and (4) sparse TSVs stiching the grounds and TSVs
terminating the resonators to the ground plane on the back. As shown in Fig.
\ref{fig:structure}(d), the sparse TSV design (2) has a spacing of 0.5 to 2 mm
between the TSVs in areas near the resonators, with each resonator having one
to three stitch vias at a distance of 150 to 300 \textmu m from the center
trace. This design aims to minimize the currents and electric fields induced
in the vias when the resonators are excited, while still providing
sufficiently dense via stitching for the ground planes to increase the
frequency of the parasitic chip modes above the measurement band. The dense
TSV design (3),
with a TSV spacing of only 0.23 mm [Fig. \ref{fig:structure}(e)],
greatly increases the participation of the stich vias in the
resonant modes by increasing their number and proximity to the center trace.
The TSV-terminated design (4) [Fig. \ref{fig:structure}(f)] makes TSVs an
integral part of the resonator, with the termination consisting of one TSV for the center conductor and three or four TSVs for the ground.
On the back side of the chip, the ends of the terminating TSVs are
approximately at a voltage node of the resonator, but the non-negligible
physical and electrical length of the vias implies that the ends of the vias
on the front side are roughly $0.02\lambda$ away from the voltage node.

\section{Quality factor measurements}

\Figure[t!](topskip=0pt, botskip=0pt, midskip=0pt){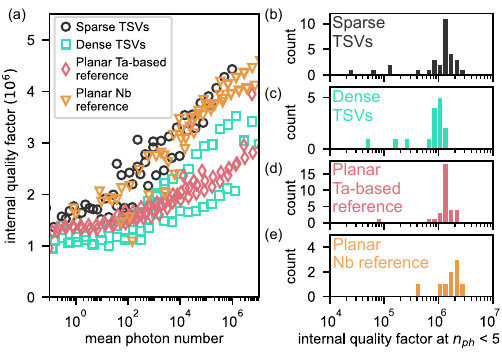}
{Quality factor measurements for via-stitched TSV chips: (a) Measured
internal quality factors $Q_{i}$ at 10 mK as a function of photon
number $n_{ph}$ circulating in the resonator for resonators with sparse
stitch TSVs (black circles), dense stitch TSVs (green squares), planar reference resonators
with the same Ta-based electrode layer (diamonds), and planar reference resonators with a Nb electrode layer (triangles). For clarity, data are shown for only two resonators with frequency between 4 GHz and 5.5 GHz for each device type. (b,c,d,e) Histogram of measured internal quality
factors at low photon numbers for all measured resonators of the types shown in panel
(a). \label{fig:2}}

Figure \ref{fig:2}(a) shows that the best resonator chips with sparse TSVs
stitching the front and back ground planes reach internal quality factors
exceeding $10^6$ at single-photon powers circulating in the resonator. The mean photon number in the resonator is nearly linearly proportional to the input probe power.
In panel (a), we show the power
dependence for a few exemplary resonators, and panels (b,c) show the
low-power $Q_{i}$ for all resonators on the measured chips. Details of the measurement setup, samples and data analysis are given in Appendices \ref{app:designs-samples} and \ref{app:measurements-analysis}.
The resonators with sparse stitch vias perform approximately identically to planar reference resonators fabricated
on either the same tantalum-based electrode layer [Fig. \ref{fig:2}(d)] or niobium [Fig. \ref{fig:2}(e)].
Furthermore, the resonator performance is similar to other reported results for silicon
substrates \cite{dunsworth_characterization_2017,nersisyan_manufacturing_2019,mcrae_materials_2020,earnest_substrate_2018,burnett_noise_2018}, and suggests that transmon-type qubits patterned on the same
electrode layer can achieve state-of-the-art coherence. This is the main
result reported in this article as it demonstrates that none of the processing
steps required to form the TSVs is fundamentally detrimental to the coherence.
Even though the TSVs are not strongly coupled to the most sensitive
long-coherence elements on the chip, the fact that via stiching of the ground
planes can be compatible with high-coherence qubits is an important
advancement in itself, as it allows physically larger quantum processor chips.

We draw this conclusion by comparing the best TSV chips to the best reference
chips, which are on par.
However, certain uniformity and yield issues remain to be solved.
This can be seen in the histogram in Fig. \ref{fig:2}(b) showing a small
fraction of outlier resonators with anomalously low quality factors, which are
relatively power independent. We occasionally find such outliers also in both Ta and Nb based 
planar reference devices. Furthermore, resonators near the edges of the 150 mm
wafers show $Q_{i}$ below $10^{5}$, even for the sparse TSV test design. In
this article, we exclude the edge chips, as wafer-level uniformity of the process has not yet been a development priority and we assume that it can be improved in the future. 

\Figure[t!](topskip=0pt, botskip=0pt, midskip=0pt){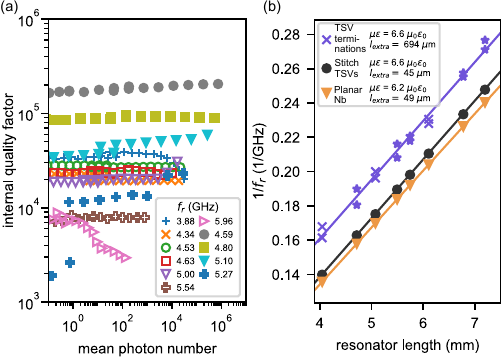} {(a) Internal
quality factor versus photon number for TSV-terminated resonators. Data points at 
high powers are not shown due to poor fit of the model to
asymmetric line shapes at high powers, and data is not shown for those
resonances where $Q_{i}\ll Q_{c}$ leading to large uncertainty in fitting. (b)
Inverse resonance frequencies of several chips with TSV-terminated resonators
(crosses and stars), resonator with sparse via stitching (circles), and reference Nb resonators (triangles) versus length of the
coplanar part. Crosses (stars) indicate a TSV termination with three (four)
ground vias around the terminating TSV. Solid lines indicate fits to Eq.
\eqref{eq:fr} with parameters given in the legend. \label{fig:tsv-terminated}}

We observe somewhat decreased $Q_{i}$ in chips with dense stich vias
{[}Fig. \ref{fig:2}(a){]}.
Furthermore, in resonators terminated with TSVs, the
internal quality factors are drastically lower, ranging from $Q_{i}$ less than
$10^{4}$ to $2\times10^{5}$, as shown in Fig.~\ref{fig:tsv-terminated}. The
line shapes of TSV-terminated resonators also generally become asymmetric at just
$10^{3}$ to $10^{5}$ photons, after which the model used to extract $Q_{i}$
from the response \cite{probst_efficient_2015} no longer fits well. In resonators 
without TSV terminations, we observe such a threshold only above
powers corresponding to over $10^{7}$ photons. Furthermore, TSV-terminated
resonators show essentially no power dependence of $Q_{i}$, until the
nonlinearity leading to asymmetric response becomes significant.

These observations suggest that, unlike in the resonators with $Q_{i}$ in the
range of a million, two-level systems (TLSs) in thin dielectric interface
layers are not a significant loss mechanism for the TSV-terminated resonators,
as they would be expected to lead to quality factors increasing with photon
number. It is instead possible that the ALD titanium nitride film on the inner
walls of the hollow TSVs contains weak spots with suppressed
superconductivity. This should lead to rapid decrease of quality factor at
powers where the current through the termination becomes comparable to the
critical current of the weak spot.
We estimate that, at the threshold power for asymmetric response,
the current through the TSV terminations is on the order of 10 microamperes (see Appendix~\ref{app:measurements-analysis}).
Such stochastically occurring weak spots could explain the large variation in $Q_i$, as well.
It is also possible that resistive losses occur at the interface between the ALD titanium nitride and the sputtered electrode layer.
Nevertheless, the best TSV-terminated resonators perform as well as
the TSV-interrupted resonators in Ref. \cite{mallek_fabrication_2021}. 

The resonance frequencies $f_{r}$ of the TSV-terminated resonators are
consistent with those of other resonators, after accounting for approximately 650 \textmu m of
CPW-equivalent length added by the TSV terminations.
The added length is qualitatively consistent with a wafer
thickness of $525 \pm 25$ \textmu m and a higher effective dielectric constant within the TSV termination, as compared to the CPW part.
To demonstrate this, Fig.~\ref{fig:tsv-terminated}(b) shows inverse resonance frequency versus
resonator length $l_{design}$ for different resonator types, as well as fits
to
\begin{equation}
1/f_{r}=4\sqrt{\mu \epsilon}(l_{design}+l_{extra}),\label{eq:fr}
\end{equation}
where the speed of light in the CPW $1/\sqrt{\mu \epsilon}$ and extra
CPW-equivalent length $l_{extra}$ are fit parameters. Neglecting kinetic inductance and film thickness \cite{ramesh_microstrip_2013}, 
we expect $\mu \epsilon = 6.23 \mu_0 \epsilon_0$ for our CPWs, 
assuming $\epsilon_r = 11.45$ for the permittivity of silicon \cite{krupka_measurements_2006}. 
For chips with stich vias only, as well as for planar reference chips,
the measured frequencies of individual resonators deviate from the linear fit by less than 0.05\%.
For TSV-terminated resonators, however, the scatter is relatively large, showing an
average deviation of 2\% even within a single chip.
The variation in the resonance frequencies from the prediction of the linear fit is not explained by variation in the termination design
[three vs four grounding TSVs surrounding the via terminating the center conductor, see Fig. \ref{fig:structure}(f)].
Figure~\ref{fig:tsv-terminated}(b) also shows that the resonance frequencies of 
resonators with stitch vias only are similar to those of planar resonators patterned on a reference Nb film.
This demonstrates the low kinetic inductance of the tantalum-based electrode layer and
therefore high compatibility with typical Nb-based designs for superconducting qubits. 

\Figure[t!](topskip=0pt, botskip=0pt, midskip=0pt){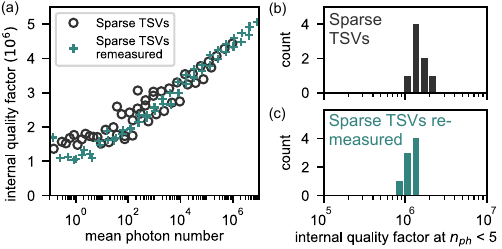}
{(a) Internal quality factor versus photon number for sparse TSV resonators
(black circles) compared to the same device measured
two weeks later (stars). For clarity, only two resonance frequencies are shown per device. 
(b,c) Histograms of low-power $Q_{i}$ for the chips shown in panel
(a). \label{fig:4}}

Figure~\ref{fig:4} shows that a chip with sparse TSV stitching continues to show internal quality factors of
$10^6$ even after the chip is left at room temperature and atmospheric pressure for two weeks after the intial measurements.
This is consistent with our observation (not shown) that planar reference
samples with the same tantalum-based electrode layer are also stable in time.

Power dependence of $Q_{i}$ is commonly used to estimate the
contribution of TLS losses, as most other loss mechanisms are expected
to be independent of power at these powers. All of the resonator types
in Fig. \ref{fig:2}(a) show relatively weak power dependence and lack
clear saturation of $Q_{i}$ at high powers, making accurate estimation
a challenge. Here, we use the common simplistic approach of defining
the total TLS loss $F\delta_{TLS}^{^{0}}$ as
$\delta_{LP}-\delta_{HP}$, where $\delta_{HP}$ ($\delta_{LP})$ is
defined as $1/Q_{i}$ at the highest (lowest) measured power.  The
obtained value $F\delta_{TLS}^{^{0}}\simeq5\times10^{^{-7}}$ is of the
same order of magnitude as best reported results for planar devices
\cite{mcrae_materials_2020}.

\section{DC characterization}

\Figure[t!](topskip=0pt, botskip=0pt, midskip=0pt){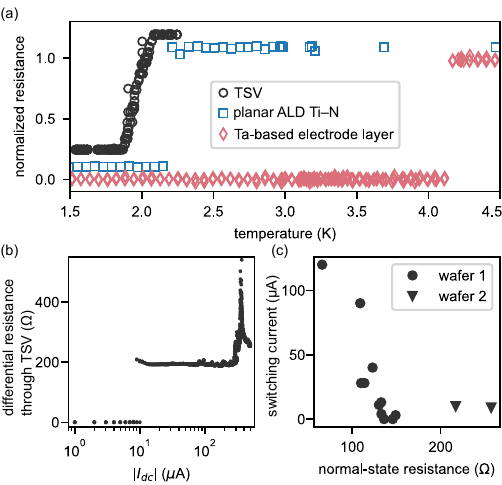}
{(a) Temperature dependence of normalized resistance for ALD titanium nitride inside a single
via (circles), planar ALD
titanium nitride (squares), and the planar tantalum-based electrode layer (diamonds).
For clarity, the traces are offset vertically in steps of 0.1.
(b) Differential resistance versus the absolute value of bias current
$I_{dc}$ through a single TSV at 100 mK, showing several switching
currents. The response is symmetric around $I_{bias}=0$. (c) Smallest
switching current versus normal-state resistance for several different
TSVs on two wafers.\label{fig:dc}}

The superconducting transition temperature of the tantalum-based electrode
layer is slightly above 4 K {[}Fig. \ref{fig:dc}(a){]}, in line with a
literature value of 4.46 K for high-purity bulk tantalum
\cite{milne_superconducting_1961}. On the planar back side of the wafer, the
ALD titanium nitride coating has a transition temperature in excess of
approximately 2 K, which is typical for highly disordered titanium nitride
deposited by ALD.
The critical temperature is significantly below highest values achieved with
ALD or other methods (4.5 to 5.4 K) \cite{torgovkin_high_2018} but easily
satisfies the basic requirement of effectively suppressing thermal
quasiparticles in superconducting qubit applications, which operate around
10~mK to 30~mK. The critical temperatures and currents were measured with
standard lock-in techniques in a four-probe configuration
(see Appendix \ref{app:measurements-analysis} for details). 

The transition of titanium nitride to the superconducting state is significantly broadened toward
lower temperatures when measured through a TSV, as shown in Fig.
\ref{fig:dc}(a). This is qualitatively similar to the broad transition
measured through TSVs lined with titanium nitride by Mallek et al. \cite{mallek_fabrication_2021}. Furthermore, Fig. \ref{fig:dc}(b) shows that the
via switches from the superconducting state to the normal state gradually, in
multiple steps from tens to hundreds of \textmu A, and the lowest switching
current varies from a few microamperes to over 100 \textmu A {[}Fig.
\ref{fig:dc}(c){]}.  Both the broad superconducting transition over temperature and observation of multiple critical currents are consistent with the
existence of weak spots in the titanium nitride lining inside the via. The
weakest spot determines the lowest switching current and, due to Joule
heating, also likely limits the highest observed switching currents to only
hundreds of \textmu A, which correspond to only a few $\text{kA}/\text{cm}^{2}$ of nominal current density. The measured critical currents are also of similar magnitude as the current through the TSVs at the threshold power where the lineshapes of TSV-terminated resonators become asymmetric (Fig. \ref{fig:tsv-terminated}). 

\section{Conclusions}

In conclusion, we successfully fabricated and characterized qubit-compatible
microwave resonators on silicon wafers with TSVs stitching the front and back
ground planes. The measured resonator internal quality factors improve over previous results
\cite{mallek_fabrication_2021} by nearly an order of magnitude and are on par
with fully planar resonator results, despite the added complexity of
fabrication. The resonator performance provides strong
evidence that state-of-the-art qubit coherence times would likely be reached
if the same process were used for transmon-type qubits, with Josephson
junctions post-processed on the samples using established evaporation-based
methods. Stitching the ground planes with TSVs is an important technique for
controlling parasitic microwave modes within the silicon chip. Without TSVs or
other methods for controlling them, the parasitic modes limit the physical
size of transmon-based quantum processor chips to the range of two centimeters.
Qubits fabricated on sapphire substrates have shown even better performance
but there is no clear path to fabricating qubit-compatible high-aspect-ratio
vias on sapphire substrates.

Critical currents of the TSVs demonstrated here leave room for future
improvement. The low switching currents in dc measurements,
the existence of outliers in the
microwave measurements, and the dramatically lower performance of
TSV-terminated resonators all hint in the direction of weak spots in the titanium
nitride film inside the vias. This may be due to roughness of the via walls or
due to imperfect conformality of the plasma-enhanced ALD process, which could
lead to variation in film quality and weaker superconductivity at the far end of
the via. 
Alternatively, the losses may be due to poor contact between the ALD titanium nitride and the sputtered titanium nitride in the electrode layer.
These potential issues can be improved without drastic
changes to the TSV structure.
Together with additional patterning of the back side metallization,
improving critical currents to the mA range would make the TSVs applicable
to flux line routing.
The low critical currents observed here limit the applications to
grounding, charge excitation lines, and readout lines.
Other possible future improvements include
increasing the aspect ratio of the vias or reducing the thickness of the
wafers, which would both lead to smaller diameter vias. Smaller diameter vias
increase integration density and would improve the mechanical stability of the
membranes covering the TSVs.

\appendices

\section{Resonator designs and samples}
\label{app:designs-samples}

\begin{table*}[t]
\begin{tabular}{p{1.4cm}|p{2cm}|p{1.0cm}|p{1cm}|p{1.4cm}|p{0.8cm}|p{0.9cm}|p{0.9cm}|p{1.2cm}|p{2.5cm}}
Design & Role of TSVs & Backside metalization & Number of resonators on chip &  $Q_c$ & Chip size (mm $\times$ mm) & Flux trapping hole diameter (\textmu m) & Flux trapping hole spacing (\textmu m) & Coupling to feedline& Note \\ \hline \hline
Sparse TSVs 1 & Stitch vias 150 to 300 µm from resonators & yes & 8  & $2 \times 10^4$ to $2 \times 10^5$  & $5 \times 7$ & 2 & 10 & capacitive & \\ 
Sparse TSVs 2 & Stitch vias 150 to 500 µm from resonators & yes & 2 & $2 \times 10^6$ to $8 \times 10^6$ & 14.3 $\times$ 14.3 & 2.5 & 8 & inductive &  \\ 
Dense TSVs & Stitch vias 100 µm from resonators & yes & 8 & $2 \times 10^4$ to $2 \times 10^5$   & 5 $\times$ 7 & 2 & 10 & capacitive & Resonator lengths and couplings identical to sparse TSVs 1 \\
TSV-terminated resonators & Stitch vias 150 to 300 µm from resonators, resonators terminate in TSVs & yes & 8 & $2 \times 10^4$ to $2 \times 10^5$  & 5 $\times$ 7 & 2 & 10 & capacitive &  Resonator lengths and couplings identical to sparse TSVs 1 \\
Reference A & None & no & 8  & $2 \times 10^4$ to $2 \times 10^5$  & 5 $\times$ 7 &  2 & 10 & capacitive & Resonator lengths and couplings identical to sparse TSVs 1  \\ 
Reference B & None & no & 8  &  $1 \times 10^4$ to $2.5 \times 10^6$ & 5 $\times$ 5 & 2.5 & 8 & inductive & \\ 
Reference C & None & no & 10 & $ 2 \times 10^5 $ to $5 \times 10^6$  & 5 $\times$ 7 & 2.5 & 8 & inductive &  \\ 
\end{tabular}
\caption{Resonator designs used in the manuscript. All resonators have a 20 \textmu m wide CPW center trace and 10 \textmu m gap between the center pin and ground electrodes. All designs incorporate a square grid of flux trapping holes. The design Reference A was only used to obtain the reference data for resonance frequencies in Nb resonators shown in Fig. 3 of the main text. \label{table:designs}}
\end{table*}

\begin{table}
\setlength{\tabcolsep}{3pt}
\begin{tabular}{c|c|p{1.7cm}|p{1.0cm}|p{3cm}}
Sample & Wafer &  Design & Sample holder & Notes \\ \hline \hline
S1 & TSV1 & Sparse TSVs 1 & Cu & \\ \hline 
S2 & TSV1 & Sparse TSVs 1 & Cu & \\ \hline 
S3 & TSV1 & Sparse TSVs 1 & Cu & \\ \hline 
S4 & TSV2 &  Sparse TSVs 2 & Cu & Base temperature 30 mK \\ \hline
S5 & TSV2 &  Sparse TSVs 2 & Cu & Base temperature 30 mK \\ \hline 
S6 & TSV2 &  Sparse TSVs 2 & Cu & Base temperature 30 mK \\ \hline 
D1 & TSV1 & Dense TSVs & Cu & \\ \hline 
D2 & TSV1 & Dense TSVs & Au/Cu & Only 10 dB attenuation at mixing chamber \\ \hline 
T1 & TSV1 & TSV-terminated resonators & Cu & \\ \hline 
T2 & TSV1 & TSV-terminated resonators & Au/Cu & Only 10 dB attenuation at mixing chamber \\ \hline 
R1 & Ta & Reference B  & Au/Cu & \\ \hline
R2 & Ta & Reference B  & Au/Cu & \\ \hline
R3 & Ta & Reference B  & Au/Cu & Base temperature 40 mK \\ \hline
R4 & Ta & Reference B  & Au/Cu & Base temperature 40 mK \\ \hline
R5 & Nb & Reference C & Cu &
\end{tabular}
\caption{Samples for which data is presented in the manuscript. Wiring is only specified if it differs from the setup shown in Fig. \ref{fig:app_wiring}. Base temperature is only specified if it exceeded 15 mK.  \label{table:samples}}
\end{table}

The results presented in this manuscript are obtained from measurements of over 100 resonators on 15 chips in several cooldowns. The measured resonators are $\lambda / 4$ coplanar waveguide resonators in a hanger-type configuration \cite{gao_physics_2008}. Transmission measurements in this configuration lead to a Lorentzian dip in the response and allow taking cable losses and impedance mismatch into account by normalizations of the measurement data with well established models \cite{khalil_analysis_2012,probst_efficient_2015, mcrae_materials_2020}. 

Each chip hosts up to 10 resonators with differing lengths. 
The resonators are either capacitively coupled to a common microwave feedline on one end and shorted to ground at the opposite one, or inductively coupled to the feedline with the opposite end left open. 
All designs have a 20 \textmu m wide CPW center trace and 10 \textmu m gap between the center pin and ground electrodes, and incorporate a square grid of flux trapping holes in the ground planes. The measured designs differ in the presence and role of TSVs and backside metalization, as well as the precise lengths of the resonators and couplings to the feedline. The design parameters along with the coupling quality factors $Q_c$ obtained from fitting the data are summarized in Table \ref{table:designs}. The resonator lengths range between 4 mm and 7 mm for all designs.

In Table \ref{table:samples}, we list the resonator chips and their measurement configurations. All resonances on all the listed chips are included in the $Q_i$ histograms shown in Figs. \ref{fig:2} of the main text. Figure \ref{fig:4} shows measurements of chip S2 in two cooldowns. We have excluded measurements performed without magnetic shielding as well as chips from the edges of the wafers from the manuscript. In Fig. \ref{fig:tsv-terminated}, we include only those resonances where the resonator fitting (described below) produced a reliable result.

\section{Measurements and data analysis}
\label{app:measurements-analysis}

\subsubsection{DC measurements}

\begin{figure}
\centering
\includegraphics[scale=0.8]{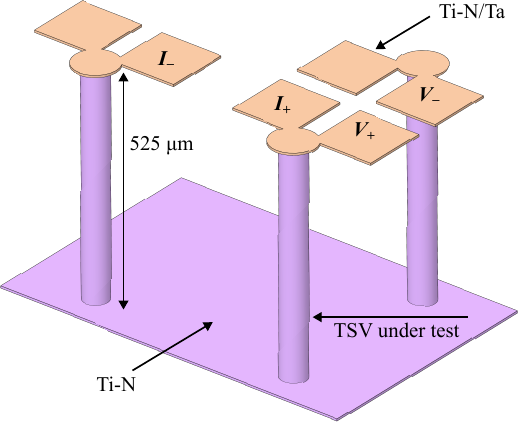} 
\caption{\label{fig:tsv-bonding} Schematic of the four-probe arrangement for measuring critical currents and temperatures through individual TSVs. Each via is connected to individual bond pads on the front side of the wafer, and the same unpatterned metalization on the back side.}
\end{figure}

For the dc characterization results shown in Fig. \ref{fig:dc}, test chips were glued to an insulating sapphire chip which was in turn mounted with vacuum grease to a copper sample holder thermally anchored to the mixing chamber stage of a dilution refrigerator. The critical temperatures of the tantalum-based electrode layer and ALD titanium nitride were measured in standard four-probe configurations with a lock-in amplifier. The measurement configuration used in the measurements of individual TSVs is schematically shown in Fig. \ref{fig:tsv-bonding}. For the critical current measurements shown in Fig. \ref{fig:dc}(c), the refrigerator temperature was stabilized to approximately 100 mK with a PID controller, as the dissipation from the TSVs in the normal state is significant compared to the cooling power of the refrigerator. 

\subsubsection{Resonator measurements\label{appendix:resonator_measurements}}

\begin{figure}[htbp]
  \centering
   \includegraphics[width=\linewidth]{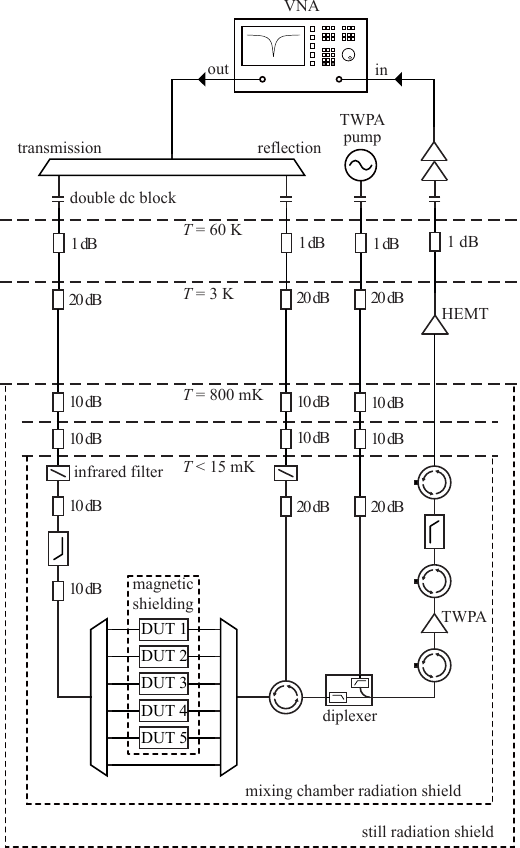}
  \caption{Measurement setup used for microwave measurements. The low-pass filters in the transmission and output lines have a cut off frequency of 12.5 GHz. The infrared filters in the transmission and return loss lines are Eccosorb filters with a low insertion loss in the 4-8 GHz band. The passband of the circulators and isolators is 4 GHz to 8 GHz or 4 GHz to 12 GHz. The control lines of the microwave switch at the mixing chamber or flux bias connections to the traveling wave parametric amplifier are not shown.  \label{fig:app_wiring}}
\end{figure}

For the resonator measurements, we wire bond chips to sample holders machined from copper or gold-plated copper thermally anchored to the mixing chamber plate of a dilution refrigerator. 
The sample holders are mounted inside magnetic shields consisting of a mu-metal shield and a superconducting aluminum tube. In most of the measurements, the magnetic shields are mounted inside a radiation shield thermalized to the mixing chamber flange, but we have not observed significant differences between samples measured inside or outside the mixing chamber shield. For most of the measurements, the base temperature of the refrigerator was below 15 mK, with exceptions indicated in Table \ref{table:samples}. 

The measurement setup used is schematically depicted in Fig. \ref{fig:app_wiring}. 
The probe signal is generated by a vector network analyser (VNA) at room temperature.
The VNA output is connected to one of two attenuated and filtered coaxial lines used for either transmission or reflection measurements. In the transmission configuration, the attenuated signal is connected to the input port of the device under test (DUT), while the reflection measurement line is connected to the other port of the 
DUT with a  circulator. 
However, in this work we only present transmission measurements.
To estimate the power reaching the DUT, 
we have measured the transmission through two identically attenuated coaxial lines connected in series at the base temperature of the refrigerator. We use this reference data as well as datasheet values for the frequency-dependent attenuation of room temperature cables and components to calculate the power reaching the DUT. 
Slight differences in the attenuation or filtering between cooldowns are indicated in Table \ref{table:samples}.

We use a pair of microwave switches at the mixing chamber to allow characterizing up to five DUTs in a single cooldown, as well as a coaxial cable that can be used as a transmission reference.
The transmitted signal is amplified with a near quantum limited three-wave mixing travelling wave parametric amplifier (TWPA) at the mixing chamber stage 
and high-electron-mobility transistor (HEMT) amplifiers at 4 K and at room temperature. 
The pump tone for the TWPA with frequency typically close to 14 GHz is combined with the signal from the DUT with a diplexer and 
filtered again from the signal after the TWPA stage to avoid saturation of the following amplifiers, while several isolators provide isolation between the DUT and the TWPA and TWPA and HEMT amplifiers, respectively. 
While the TWPA decreases the measurement time required for accurate characterization at single-photon powers circulating in the resonators, the largest probe signals would saturate it. We thus turn the pump tone off at high probe powers and verify that the internal quality factors extracted at intermediate powers are the same with the pump tone on and off.

\subsubsection{Extraction of the internal quality factor}

In this manuscript, we have used the open source fitting routine of Ref. \cite{probst_efficient_2015} to extract the quality factors from the measured data.
For resonators in the hanger configuration, the model for the transmitted signal $S_{21}(f)$ reads 
\begin{equation}
S_{21}(f) = a\mathrm{e}^{{\mathrm{i}\alpha}}\mathrm{e}^{-2\pi \mathrm{i} f \tau}\left[ 1-\frac{\frac{Q}{\left| Q_{e}\right|}\mathrm{e}^{\mathrm{i}\phi}}{1+2\mathrm{i}Q\left( \frac{f}{f_r}-1\right)} \right]\text{.}
\label{eq:reso_model}
\end{equation}
The term in front of the brackets covers contributions from the measurement environment where $a$ describes the baseline level of transmission, $\alpha$ the phase shift and $\tau$ the electrical delay across the measurement line. $f$ is the probe and $f_\mathrm{r}$ the resonance frequency. The model is based on the diameter correction method (DCM) of Ref. \cite{khalil_analysis_2012} where the complex coupling quality factor
\begin{equation}
Q_{e} = \left|Q_e\right| \mathrm{e}^{-\mathrm{i}\phi}
\end{equation}
with magnitude $|Q_e|$ and rotation angle $-\phi$
accounts for asymmetries in the Lorentzian shape, e.g., due to impedance mismatch between the resonator and feedline, or the feedline and the measurement environment \cite{khalil_analysis_2012, deng_analysis_2013}. 
The loaded quality factor $Q$ is then given by
\begin{equation}
Q^{-1} = Q_i^{-1} + Q_c^{-1}
\end{equation}
where 
\begin{equation}
Q_c^{-1} = \Re\left\{Q_e^{-1}\right\}
\end{equation}
and $Q_i$ is the internal quality factor.

Following the circuit analysis of Refs. \cite{barends_photondetecting_2009, devisser_quasiparticle_2014}, the root mean square voltage $V_r$ of the standing wave inside a hanger type CPW resonator at resonance is given by
\begin{equation}
V_r = Q\sqrt{\frac{2Z_r}{\pi Q_c}\left(\frac{l}{\lambda}\right)^{-1}P_{dev}}\label{eq:voltage_resonator}
\end{equation} 
with the characteristic impedance $Z_r$ of the resonator's CPW. $\lambda$ is the wavelength at resonance, $l$ is the length of the resonator and $P_{dev}$ is the power entering the DUT. This equation is valid for both quarter- and half-wave resonators and their higher frequency modes as well. The average energy $E$ inside such CPW resonator is given by
\begin{equation}
E = \frac{1}{2 f_r Z_r} \frac{l}{\lambda} V_r^2 \text{.} \label{eq:energy_resonator}
\end{equation}
Thus with Eqs. \eqref{eq:voltage_resonator} and \eqref{eq:energy_resonator}, we calculate the average number of photons circulating in the resonator in accordance with Ref. \cite{bruno_reducing_2015} as
\begin{equation}
	n_{ph}=\frac{E}{hf_r} = \frac{1}{\pi}\frac{Q^2}{Q_c}\frac{P_{dev}}{h f_r^2} \label{eq:photon_number}
\end{equation}
where $h$ is Planck's constant. 
We estimate $P_{dev}$ from the output power at the room-temperature generator as described in section \ref{appendix:resonator_measurements} above. 
Note that Eq. \eqref{eq:photon_number} is valid for any resonator in hanger configuration. However, in a one-port reflection measurement, the right hand side of the equation would be multiplied by a factor of $2$.
In the $Q_i$ histograms of Figs. \ref{fig:2} and \ref{fig:4} of the main text, we show for each resonator the mean $Q_i$ from all measurements with $n_{ph} < 5$.

The TSV terminations of the devices shown in Fig. \ref{fig:tsv-terminated} of the main text represent the shorted end of the quarter-wave resonators and are therefore located at the current maxima of the standing waves. 
Thus for $l=\lambda/4$, the maximum current flowing through the TSV terminations can be estimated from 
\begin{equation}
\hat{I}_{TSV} = \sqrt{2} \frac{V_r}{Z_r} = 4Q\sqrt{\frac{P_{dev}}{\pi Q_c Z_r}}
\end{equation}
with $Z_r = 50$~$\Omega$.

\section*{Acknowledgment}

We acknowledge 
Jan Toivonen,
Harri Pohjonen, Ville Selinmaa, 
Paula Holmlund, and Jaana Marles for technical assistance.

This work was financially supported by OpenSuperQ, which has received
funding from the European Union's Horizon 2020 research and innovation
programme under grant agreement no. 820363. 
The work at VTT was supported 
by the Quantum Computer Codevelopment project funded by the Finnish government, 
and performed as part of the Academy of Finland Centre of Excellence program (projects 336817, 312059 and 312294).
We also acknowledge financial support from the European Commission H2020 project EFINED (grant agreement no. 766853).
The Chalmers work was in part supported by the Wallenberg Center for Quantum Technology, and performed at Myfab Chalmers.
V. Vesterinen acknowledges financial support from the Academy of Finland through Grant No. 321700.

\section*{Author contributions}

K. Grigoras led process development of via etching and ALD and was
a key contributor to electrode and membrane layer development. 
N. Yurttag{\"u}l and J.-P. Kaikkonen led process development of the electrode and membrane layers. 
E. T. Mannila led characterization and analysis efforts together with N. Tiencken. 
P. Eskelinen led process development of the final ALD recipe.  
A. Ronzani, J. H{\"a}tinen and S. Simbierowicz contributed to dc characterization throughout the process development. 
S. Simbierowicz also contributed to mask design.  
L. Gr{\"o}nberg, V. Vesterinen and S. Simbierowicz contributed to development of relevant planar resonators.
D. Datta and V. Vesterinen contributed to rf measurements.
D. P. Lozano did process development of electrode layer deposition and device characterization at Chalmers, used in earlier iterations of the devices that the reported devices improve upon. 
H.-X. Li contributed to device characterization. M. Rommel contributed to process development. D. Shiri contributed to microwave design and analysis.
J. Bizn{\'a}rov{\'a}, A. Fadavi Roudsari, S. Kosen, and A. Osman assisted in fabrication, characterization, and discussions. 
J. Bylander supervised the work at Chalmers.  
J. Govenius supervised and contributed technical advice to all aspects of the work, 
after the first phases supervised and planned by M. Prunnila and J. Hassel.



\EOD

\end{document}